\documentclass[11pt]{article}
\usepackage{graphicx}
\usepackage{epsfig} \usepackage{amssymb} \usepackage{amsfonts}
\def\agoth{\relax\ifmmode{\mathfrak A}\else{${\mathfrak A}${ }}\fi}
\def\pisq{\relax\ifmmode{\pi^2}\else{${\pi^2}${ }}\fi}
\def\agothk{\relax\ifmmode{\mathfrak A}_k\else{${\mathfrak A}_k${ }}\fi}
\def\acal{\relax\ifmmode{\cal A}\else{${\cal A}${ }}\fi}
\def\acalk{\relax\ifmmode{\cal A}_k\else{${\cal A}_k${ }}\fi}
\newcommand{\beq}{\begin{equation}} \newcommand{\eeq}{\end{equation}}
\newcommand{\be}{\begin{equation}}
\newcommand{\ee}{\end{equation}}
\newcommand{\bea}{\begin{eqnarray}}
\newcommand{\eea}{\end{eqnarray}}
\begin{document}
\begin{center}
{\large\bf Practical techniques of analytic perturbation theory of QCD}
\medskip

{\bf B.A. Magradze \footnote{Tbilisi Mathematical Institute, 380093 Tbilisi, Georgia, E-mail: magr@rmi.acnet.ge.}

}

\medskip

{\footnotesize \bf Abstract}
\medskip

\parbox{110mm}{\footnotesize The Lambert-W explicit solutions to the QCD renormalization group (RG) equation are considered up to fourth order in the ${\overline {MS}}$ scheme.
We compare, systematically, these solutions with the conventional asymptotical (iterative) approximations and with the exact numerical solutions to the RG equation in the domain with three quark flavours. Applications of these solutions in analytic perturbation theory (APT) are discussed. Using these (Lambert-W, asymptotical and exact numerical) solutions we reconstruct the expansion functions for the non-power APT series in the space- and time-like regions.
These expansion functions are examined in the infrared region. It is shown that the Lambert-W solutions provide the excellent accuracy.}
\end{center}
\medskip
\section{Introduction}
Recently, in works \cite{my,ggk} the exact explicit 2-loop solution to the QCD  renormalization group (RG) equation was considered. This solution is given by
\be
\label{w2}
\alpha_{s}^{(2)}(Q^2,f)=-\frac{\beta_0}{\beta_1}\frac{1}{1+
W_{-1}(\zeta)},\quad\mbox{where}\quad
\zeta=-\frac{1}{eb_{1}}\left(\frac{Q^2}{\Lambda^2}\right)^{-\frac{1}{b_{1}}},
\ee
$f$ denotes the number of active quark flavours at the energy scale $Q$, $\beta_{k}$ is the kth order coefficient of the QCD $\beta$-function, $b_{1}=\beta_1/\beta_0^2$,
 $\Lambda\equiv\Lambda_{\overline {MS}}$ is the conventional QCD parameter and $W(\zeta)$ is the Lambert W
function.
The branches of W are denoted $W_{k}(\zeta), k=0,\pm 1,\ldots .$ An exhaustive review of the Lambert-W function may be found in Ref.~\cite{lamb}.
In our notations the RG equation has the form
\footnote{Here $Q^2=-q^2$ and $Q^2>0$ in the Euclidean domain.}
\be
\label{eff}
Q^2 \frac{\partial{\alpha_{s}(Q^2,f)}}{\partial{Q^2}}=
{\beta}^{f}(\alpha_{s}(Q^2,f))=-\sum_{n=0}^{\infty}
\beta_{n}^{f}\alpha_{s}^{n+2}(Q^2,f),
\ee
the normalization condition is $\alpha_{s}(\mu^{2},f)=g^2/(4\pi)$, where $\mu$ is the renormalization point and $g$ is the gauge coupling of QCD.
In the class of schemes where the beta-function is mass
independent $\beta_0^{f}$ and $\beta_1^{f}$ are universal
\be
\label{coef}
\beta_{0}^{f}=\frac{1}{4\pi}\left(11-\frac{2}{3}f\right),\qquad
\beta_{1}^{f}=\frac{1}{(4\pi)^2}\left(102-\frac{38}{3}f\right),
\ee
the results for $\beta_{2}^{f}$  \cite{tvz} and $\beta_{3}^{f}$ \cite{rvl} in the modified $MS$ $(\overline {MS})$ scheme are
\be
\beta_{2}^{f}=\frac{1}{(4\pi)^3}\left(\frac{2857}{2}-\frac{5033f}{18}+\frac{325f^2}{54}\right),
\ee
\be
\beta_{3}^{f}=\frac{1}{(4\pi)^4}\left(\frac{149753}{6}+3564\zeta_{3}-\left(\frac{1078361}{162}+\frac{6508}{27}\zeta_{3}\right)f+
\left(\frac{50065}{162}+\frac{6472}{81}\zeta_{3}\right)f^2+\frac{1093}{729}f^3\right).
\ee
Here $\zeta$ is the Riemann zeta-function ($\zeta_{3}=1.202056903...$).

Using (\ref{w2}) the analytical continuation of the coupling in the complex $Q^2$ plane has been performed \cite{my,ggk,my1} (see also papers \cite{my2,mypr}).
Afterwards, in paper \cite{kur}, a higher order running coupling
(in arbitrary $\overline{MS}$-like renormalization scheme) was expanded
in powers of the scheme independent explicit two-loop coupling (\ref{w2})
\footnote{here and hereof we omit the argument $f$.}
\be
\label{ser}
\alpha_{s}^{(k)}(Q^2)=\sum_{n=1}^{\infty}c_{n}^{(k)}\alpha_{s}^{(2)n}(Q^2),
\ee
then, a new method for reducing the scheme ambiguity for the QCD observables was proposed. A similar expansion  (motivated differently for an observable depending on a single scale) has been suggested in \cite{mx}.

The exact solution (\ref{w2}) and the approximation (\ref{ser}) proved to be useful in the dispersive (analytic) approach to perturbative QCD.
The dispersive approach \cite{ss,ss0, dmw,gr,mss,cane} has been devised to extend the appropriately modified perturbation theory calculations towards the low-energy frontier.
In work \cite{mss} a particular version of the analytic approach, the analytic
perturbation theory (APT), was suggested.
According to this approach the standard perturbative series in powers of the running coupling $\alpha_{s}(Q^2)$ are replaced by a novel, non-power series.
The individual terms of these series (determined by the functions $\acal_{n}(Q^2)$) provide the rigor momentum-plane analyticity properties for the expanded quantity itself.
APT has several (phenomenological) advantages over the conventional perturbation theory even for moderate energies in the five-flavour region \cite{sh3,sh4, sh5}. A striking feature of the non-power expansions in APT is that they exhibit improved convergence properties as compared with the standard perturbation theory \cite{mss, sh4, sh5}. Thus, the large $\pi^2$-contributions to the expansion coefficients of time-like observables, inherent in standard perturbation theory \cite{rad,krp}, are automatically summed in the individual terms of the non-power series. For various phenomenological and theoretical consequences obtained within APT we refer to works \cite{sh4,ss1}.

In works \cite{mypr} and \cite{join} the Euclidean and Minkowskian sets of functions \footnote{i.e. the functions of the Euclidean and Minkowskian arguments.}
$\{\acal_n(Q^2)\}$ and $\{\agoth_{n}(s)\}$ have been explicitly constructed. The Lambert-W solutions (\ref{w2}) and (\ref{ser}) were used. The results, up to third order, were given in a compact form. Thus, the integrals for the time-like set of functions (see below formula (\ref{eleg})) were analytically performed. For both sets of the functions an elegant equations have been derived; in the 2-loop case \cite{mypr} and afterwards, also, in all orders of perturbation theory \cite{join}. These equations generalize the one loop equations obtained previously in works \cite{sh5}. Thus functions $\acal_{n}^{(k)}$ satisfy
the equations
\be
\label{rec4}
\frac{\partial \acal_{n}^{(k)}(Q^2)}{\partial \ln Q^2}=
-n\sum_{N=0}^{k-1}\beta_{N}^{f}\acal_{n+N+1}^{(k)}(Q^2),\qquad n=1,2\ldots.
\ee
Global variants of these functions with the quark threshold effects  have been also reconstructed \cite{sh3,sh4,mypr,join}.

In this paper we estimate accurateness of the explicit Lambert-W solutions
(\ref{ser}) and the conventional asymptotical approximations (see below formula (\ref{as})) at low scales. We keep in mind applications of these solutions in conventional and analytic perturbation theories. As a standard for comparison the ``exact'' numerically determined running coupling is used. In sect. 2, we present relevant formulas for the running coupling up to 4-loop order and give numerical results obtained from the approximate and exact solutions. In sect. 3, we present theoretical formulas for the spacelike  and timelike  ``analytic'' images of powers of the running coupling (the functions $\acal_{n}(Q^2)$ and $\agoth_{n}(s)$). To reconstruct these functions, the Lambert-W as well as the asymptotical solutions are applied. In the infrared region we give numerical results obtained from these approximations to the images, together with the ``exact'' results obtained by using a numerical method. Sect. 4 contains conclusions.

\section{The Lambert-W solutions vs the asymptotical solutions}

In practice, it is sufficient to retain the first few terms in the expansion (\ref{ser}).
Then the approximation to the coupling reads
\be
\label{trs}
\alpha_{s,ts}^{(k)}(Q^2)=\sum_{n=1}^{n=N_{tr}}c_{n}^{(k)}\alpha_{s}^{(2)n}(Q^2)
\ee
where $k=3,4\ldots$ and $c_{1}^{(k)}=1$. The coefficient $c_{2}^{(k)}$ is arbitrary; this reflects the arbitrariness in the definition of the $\Lambda$ parameter. The conventional ${\overline {MS}}$ definition of $\Lambda$ \cite{bbdm} corresponds to $c_{2}^{(k)}=0$.  The coefficients $c_{n}^{(k)}$, for $n>2$, may be readily determined. In the 4-loop order, the first nine coefficients are
\be
\label{coeffs}
c_{3}^{(4)}=\frac{\beta_{2}}{\beta_{0}},\quad c_{4}^{(4)}=\frac{\beta_{3}}{2\beta_{0}},\quad c_{5}^{(4)}=\frac{5}{3}\frac{\beta_{2}^2}{\beta_{0}^2}-\frac{\beta_{1}\beta_{3}}{6\beta_{0}^2},\ee
$$c_{6}^{(4)}=-\frac{1}{12}\frac{\beta_{1}{\beta_{2}}^2}{\beta_{0}^3}+\frac{1}{12}\frac{\beta_{3}{\beta_{1}}^2}{\beta_{0}^3}+2\frac{\beta_{2}{\beta_{3}}}{\beta_{0}^2},$$
$$c_{7}^{(4)}=\frac{1}{20}\frac{\beta_{1}^2\beta_{2}^2}{\beta_{0}^4}+
\frac{16}{5}\frac{\beta_{2}^3}{\beta_{0}^3}-\frac{4}{5}\frac{\beta_{1}\beta_{2}\beta_{3}}{\beta_{0}^3}+\frac{11}{20}\frac{\beta_{3}^2}{\beta_{0}^2}-\frac{1}{20}\frac{\beta_{3}\beta_{1}^3}{\beta_{0}^4},$$
$$c_{8}^{(4)}=\frac{19}{3}\frac{\beta_{3}\beta_{2}^2}{\beta_{0}^3}-\frac{23}{60}\frac{\beta_{1}\beta_{2}^3}{\beta_{0}^4}+\frac{9}{20}\frac{\beta_{2}\beta_{3}\beta_{1}^2}{\beta_{0}^4}-\frac{49}{120}\frac{\beta_{1}\beta_{3}^2}{\beta_{0}^3}-\frac{1}{30}\frac{\beta_{1}^3\beta_{2}^2}{\beta_{0}^5}+\frac{1}{30}\frac{\beta_{3}\beta_{1}^4}{\beta_{0}^5}.$$
$$
c_{9}^{(4)}=-\frac{946}{315}\frac{\beta_{1}\beta_{2}^2\beta_{3}}{\beta_{0}^4}
-\frac{41}{140}\frac{\beta_{1}^{3}\beta_{2}\beta_{3}}{\beta_{0}^5}
+\frac{134}{35}\frac{\beta_{2}\beta_{3}^2}{\beta_{0}^3}
+\frac{2069}{315}\frac{\beta_{2}^4}{\beta_{0}^4}
+\frac{103}{420}\frac{\beta_{1}^{2}\beta_{2}^3}{\beta_{0}^5}
+\frac{149}{504}\frac{\beta_{1}^{2}\beta_{3}^2}{\beta_{0}^4}
+\frac{1}{42}\frac{\beta_{1}^{4}\beta_{2}^2}{\beta_{0}^6}
-\frac{1}{42}\frac{\beta_{1}^{5}\beta_{3}}{\beta_{0}^6}.
$$
The coefficients have been calculated with symbolic manipulation system Maple V (release 5). It is convenient to apply the system Maple V to carry out numerical calculations, when using analytical expressions for $\acal_{n}(Q^2)$ and $\agoth_{n}(s)$ determined in terms of the Lambert W-function (see below).
Maple has an arbitrary precision implementation of all branches of the Lambert-W function. On the other hand, for numerical problems professional FORTRAN compiler is preferable. This allows one to reduce the necessary computer time. However, the Lambert-W function is not available in standard FORTRAN libraries.
To avoid this obstacle,
one may numerically integrate the RG equation in the complex $Q^2$-plane, instead of using the Lambert-W solutions. To the kth order the implicit solution to the RG equation is given by \cite{stev}
\be
\label{gsol}
\ln \left(\frac{Q^2}{{\Lambda_{\overline {MS}}}^2}\right)=\frac{1}{a_{s}}-b_{1}\ln\left(b_{1}+\frac{1}{a_{s}}\right)+\int_{0}^{a_{s}}
\left(\frac{1}{{\overline{\beta}}^{(k)}(x)}-\frac{1}{{\overline{\beta}}^{(2)}(x)}\right)dx,
\ee
where we have introduced the quantities $a_{s}(\frac{Q^2}{{\Lambda_{\overline {MS}}}^2})=\beta_{0}\alpha_{s}(Q^2)$ and ${\overline{\beta}}^{(k)}(a_{s})=\beta_{0}{\beta}^{(k)}(\frac{a_{s}}{\beta_{0}})$, and $b_{1}=\frac{\beta_{1}}{\beta_{0}^2}$. In the 3- and 4-loop cases the integral on the right (\ref{gsol}) can be done. In the 3-loop case we find
\be
\label{3lim}
\ln \left(\frac{Q^2}{{\Lambda_{\overline {MS}}}^2}\right)=\frac{1}{a_{s}}-\frac{b_1}{2}\ln \left(\frac{1}{a_{s}^{2}}+\frac{b_1}{a_s}+b_2\right)+\frac{2b_{2}-b_{1}^2}{\sqrt{\Delta^{(3)}}}
\left(\arctan\left(\frac{b_{1}+2b_{2}a_{s}}{\sqrt{\Delta^{(3)}}}\right)-
\arctan\left(\frac{b_{1}}{\sqrt{\Delta^{(3)}}}\right)\right),
\ee
where $b_{k}=\beta_{k}/\beta_{0}^{k+1}$ and $\Delta^{(3)}=4b_{2}-b_{1}^2$. Formula (\ref{3lim}) is valid only for $\Delta^{(3)}>0$ (in the ${\overline{MS}}$ scheme, for $f=3$, $\Delta^{(3)}\approx 2.91$).
In the 4-loop case the result is more complicated
\be
\label{4lim}
\begin{array}{l}
\ln \left(\frac{Q^2}{{\Lambda_{\overline {MS}}}^2}\right)=\displaystyle{\frac{1}{a_{s}}}-b_{1}\ln\left(b_{1}+\displaystyle{\frac{1}{a_{s}}}\right)+\displaystyle{\frac{1}{b_{3}(1+b_{1}x_{1})}J^{(4)}(a_{s})}\\
J^{(4)}(a_{s})=\displaystyle{\frac{b_{2}+b_{3}x_{1}}{2{\cal R}(x_{1})}\ln \left(\frac{(a_{s}-x_{1})^{2}n}{x_{1}^{2}{\cal R}(a_{s})}\right)}-\frac{b_{2}-b_{3}/b_{1}}{2{\cal R}(-1/b_{1})}\ln \left(\frac{(a_{s}+1/b_{1})^{2}b_{1}^{2}n}{{\cal R}(a_{s})}\right)\\
-\displaystyle{\left(\frac{b_{2}+b_{3}x_{1}}{{\cal R}(x_{1})}(m+2x_{1})-\frac{b_{2}-b_{3}/b_{1}}{{\cal R}(-1/b_{1})}\left(m-\frac{2}{b_{1}}\right)\right)\frac{1}{\sqrt{\Delta^{(4)}}}\left(\arctan\left(\frac{2a_{s}+m}{\sqrt{\Delta^{(4)}}}\right)-\arctan\left(\frac{m}{\sqrt{\Delta^{(4)}}}\right)\right)}
\end{array}
\ee
where we denote
$$
\begin{array}{l}
\displaystyle{m=u+v+\frac{2}{3}r},\quad n=\frac{1}{4}m^2+\frac{3}{4}(u-v)^2,\quad \Delta^{(4)}=3(u-v)^2,\quad x_{1}=u+v-\frac{1}{3}r,\\
\displaystyle{{\cal R}(x)=x^2+m x +n,}\quad u=(\sqrt{\delta}-q/2)^{1/3},\quad v=-(\sqrt{\delta}+q/2)^{1/3},\\
\displaystyle{\delta=(p/3)^3+(q/2)^2},\quad p=s-\frac{1}{3}r^2,\quad q=\frac{2}{27}r^3-\frac{1}{3}r s+\frac{1}{b_{3}},\quad r=\frac{b_{2}}{b_{3}}, s=\frac{b_{1}}{b_{3}}, b_{k}=\frac{\beta_{k}}{\beta_{0}^{k+1}}.
\end{array}
$$
Formula (\ref{4lim}) has been obtained under the assumptions $\delta>0$ and $\sqrt{\delta}>q/2$ (in the ${\overline {MS}}$ scheme, for $f=3$, $\sqrt{\delta}-q/2\approx 0.003$).

In work \cite{msy}, the 3-loop Eq.~(\ref{3lim}) has been numerically solved in the complex $Q^2$ plane. In this way the spectral densities (discontinuities) were numerically calculated. Evidently, from the technical point of view, this approach is more complicated than the using of the Lambert-W solutions.
Strictly speaking, mathematical conditions for using of the numerical method should be investigated, i.e. the analytical structure of the coupling must be established. In a mathematically rigor manner this task has been performed only in the 2-loop case \cite{my,ggk,my1}. Nevertheless, in this article, we shall use this method as well, up to fourth order in APT.

In QCD, practitioners use asymptotical (iterative) solution to the RG equation. It is presented in the form of the asymtotical series \cite{btk}
\be
\label{as}
\begin{array}{l}
\alpha_{s,as}(Q^2)=\displaystyle{\frac{1}{\beta_0
L}-\frac{\beta_1}{\beta_0^3}\frac{\ln L}{L^2}} \\
\qquad +\displaystyle{\frac{1}{{\beta}_{0}^{3}
L^3}\left(\frac{\beta_1^2}{\beta_0^2}(\ln^2 L-\ln
L-1)+\frac{\beta_2}{\beta_0}\right)}\\
\qquad{}+\displaystyle{\frac{1}{\beta_{0}^{4}L^4}\left(\frac{\beta_{1}^3}{\beta_{0}^3}\left(-\ln^{3}L+\frac{5}{2}\ln^{2}L+2\ln L-\frac{1}{2}\right)-3\frac{\beta_1\beta_2}{\beta_0^{2}}\ln L+\frac{\beta_3}{2\beta_0}\right)+O\left(\frac{\ln^4 L}{L^5}\right)},
\end{array}\ee
where $L=\ln Q^2/{\Lambda}^{2}_{\overline MS}$. The first line of Eq.~(\ref{as}) includes the 1- and 2-loop contributions to the coupling, the second line is the 3-loop correction and the third line is 4-loop correction. In the 3-loop case,  coupling (\ref{as}) has been frequently used in APT for calculation several spacelike and timelike observables \cite{msy,gi}. On the other hand, for small momentum transfer the approximant (\ref{as}) differs, significantly, from the exact numerical solution to the RG equation (see Table 1). In fact, formula (\ref{as}) is valid for $L\gg 1$. Therefore, application of (\ref{as}), within infrared region, requires a justification.

Throughout in this paper we consider QCD with $f=3$ active flavours. For the QCD parameter we take the reference value  $\Lambda_{\overline{MS}}=0.400$ $GeV$ uniformly for all approximants considered. One could use a more sophisticated method to compare the approximants, but this could have caused small changes of the final results.

In Table 1, we compare
the Lambert-W solution (\ref{trs}) to third order (with $N_{tr}=5$) \footnote{The 3-loop order coefficients $c_{n}^{(3)}$ can be read from (\ref{coeffs}) setting $\beta_3=0$.}
and the asymptotical solution (\ref{as}) (to second and third orders) with the 3-loop ``exact'' numerical coupling $\alpha_{num}^{(3)}$,  the numerical solution of Eq.~(\ref{3lim}).
Diff.(\%,) stands for the percentage deviations: Diff.(\%,ts.3)$=(1-\alpha_{ts}^{(3)}/\alpha_{num}^{(3)})*100$, Diff.(\%,as.k)$=(1-\alpha_{as}^{(k)}/\alpha_{num}^{(3)})*100$, $k=2,3$. The best accuracy is achieved with the Lambert-W approximant $\alpha_{ts}^{(3)}$.

In Table 2, we summarize the results obtained from the
3- and 4-loop couplings. $\alpha_{num}^{(4)}$ denotes the ``exact'' numerical solution of Eq.~(\ref{gsol}) in the 4-loop case (see Eq.~(\ref{4lim})), $\alpha_{ts}^{(4)}$ is the 4-loop Lambert-W solution (\ref{trs}) for $N_{tr}=9$ and $\alpha_{as}^{(k)}$ (where k=3,4) refers to
the asymptotical coupling (\ref{as}) at the 3- and 4-loop orders.
We see again, that the best accuracy achieved with $\alpha_{ts}^{(4)}$.
Curiously, in this region, the 3-loop asymptotical coupling $\alpha_{as}^{(3)}$ gives a better approximation to the exact 4-loop coupling than $\alpha_{as}^{(4)}$.

In Table 3 we give the percentage deviations of the Lambert-W approximations (\ref{trs}) to fourth order (for $N_{tr}=$4,6,9) from $\alpha_{num}^{(4)}$.
We see, that the approximation (\ref{trs}) is gradually improved when $N_{tr}$ increases. The convergence of the series (\ref{ser}) worsens only close to the Landau singularity (in the 4-loop case the Landau singularity occurs at $Q_{L}^{(4)}\approx 1.773\Lambda_{\overline{MS}}\approx 0.709$ $GeV$ for $f=3$ and $\Lambda_{\overline{MS}}=0.400$ $GeV$). In Figure 1 we plot the ``exact'' numerical coupling $\alpha_{num}^{(4)}(Q^2)$ and the approximants $\alpha_{ts}^{(4)}(Q^2,N_{tr})$ (for $N_{tr}=4,6, 9$) as a function of the momentum variable $Q=\sqrt{Q^2}$.
\begin{figure}
\centering
\resizebox{15 cm}{10cm }{\includegraphics{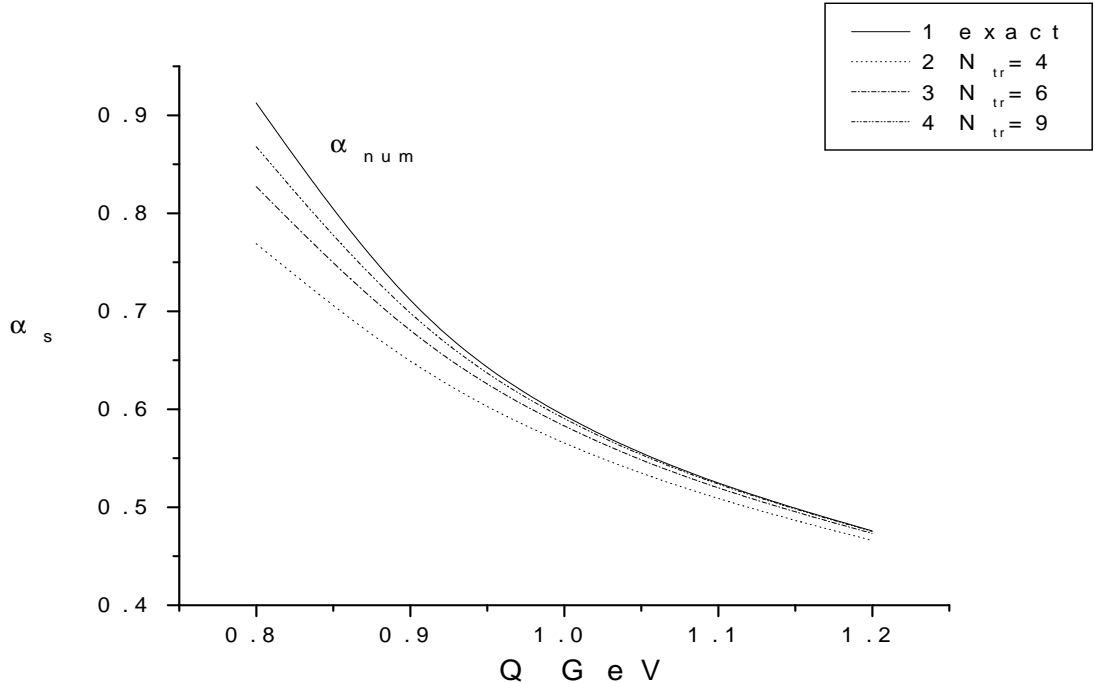}}
\caption{Plot of the functions $\alpha_{num}^{(4)}(Q^2)$ and $\alpha_{ts}^{(4)}(Q^2,N_{tr})$ ( $N_{tr}=4,6,9$). The solid line refers to the ``exact'' numerical coupling $\alpha_{num}^{(4)}(Q^2)$.}
\end{figure}

\begin{center}
\begin{table}[h]
\caption{Percentage deviations of the approximants $\alpha_{ts}^{(3)}(N_{tr}=5)$, $\alpha_{as}^{(3)}$ and $\alpha_{as}^{(2)}$ from the exact (numerical) 3-loop coupling $\alpha^{(3)}_{num}$.}
\vspace{0.2cm}
\begin{center}

\begin{tabular}{lllll}
\hline
$\sqrt{Q^{2}}$  $ GeV $   &
$ \alpha_{num}^{(3)}$ &
Diff.(\%,ts.3)&
Diff.(\%,as.3)&
Diff.(\%,as.2) \\ \hline

.80   &  0.76491 &   1.87  & -15.49 & -7.16  \\
.90   &  0.63323 &   0.81  & -9.25  & -3.93 \\
1.00  &  0.55414 &   0.43  & -6.08  & -1.59  \\
1.10  &  0.50028 &   0.26  & -4.33  &  0.01   \\
1.20  &  0.46075 &   0.17  & -3.29  &  1.12  \\
1.30  &  0.43025 &   0.12  & -2.62  &  1.90  \\
1.40  &  0.40587 &   0.09  & -2.17  &  2.47   \\
1.50  &  0.38583 &   0.06  & -1.86  &  2.88  \\
1.60  &  0.36901 &   0.05  & -1.63  &  3.19   \\
1.80  &  0.34220 &   0.03  & -1.33  &  3.60    \\
2.00  &  0.32165 &   0.02  & -1.14  &  3.84    \\
2.2   &  0.30527&   0.02  & -1.01  &  3.98     \\
2.6   &  0.28059&   0.01  & -.85   &  4.11      \\

\hline
\end{tabular}
\end{center}
\end{table}
\end{center}

\begin{center}
\begin{table}[h]
\caption{The approximants $\alpha_{ts}^{(4)}(N_{tr}=9)$, $\alpha_{as}^{(4)}$ and $\alpha_{as}^{(3)}$ vs the 4-loop``exact'' numerical coupling $\alpha_{num}^{(4)}$. Percentage deviations of these approximants from
$\alpha_{num}^{(4)}$ are given.}
\vspace{0.2cm}
\begin{center}

\begin{tabular}{llllllll}
$\sqrt{Q^{2}}$  $ GeV $   &
$ \alpha_{num}^{(4)}$ &
$ \alpha_{ts}^{(4)}$ &
$ \alpha_{as}^{(4)}$ &
$ \alpha_{as}^{(3)}$ &
Diff.(\%,ts)&
Diff.(\%,as4)&
Diff.(\%,as3)
\\ \hline

0.80 & 0.91262 & 0.86804 & 1.04504 &  0.88340  & 4.89  &  -14.5  & 3.20  \\
0.90 & 0.69243 & 0.68439 & 0.76687 &  0.69179  & 1.16  &  -10.7  & 0.09  \\
1.00 & 0.58701 & 0.58462 & 0.63081 &  0.58784  & 0.41  &  -7.46  &  -0.14  \\
1.10 & 0.52156 & 0.52063 & 0.54982 &  0.52195  &  0.18  & -5.42  &  -0.07  \\
1.20 & 0.47585 & 0.47542 & 0.49548 &  0.47589  &  0.09  & -4.12  &  -0.01  \\
1.30 & 0.44164 & 0.44142 & 0.45606 &  0.44153  &  0.05  & -3.26  &  0.03  \\
1.40 & 0.41484 & 0.41471 & 0.42589 &  0.41469  &  0.03  & -2.66  &  0.04 \\
1.50 & 0.39313 & 0.39305 & 0.40190 &  0.39301  &  0.02  & -2.23  &  0.03  \\
1.60 & 0.37510 & 0.37505 & 0.38224 &  0.37503  &  0.01  & -1.90  &  0.02   \\
1.70 & 0.35984 & 0.35980 & 0.36577 &  0.35982  &  0.01  & -1.65  &  0.00   \\
1.80 & 0.34670 & 0.34667 & 0.35173 &  0.34674  &  0.01  & -1.45  &   -0.01 \\
2.00 & 0.32515 & 0.32514 & 0.32892 &  0.32530  &  0.00  & -1.16  &   -0.05 \\
2.20 & 0.30811 & 0.30810 & 0.31106 &  0.30835  &  0.00  & -0.96   &  -0.08 \\
2.40 & 0.29421 & 0.29421 & 0.29660 &  0.29452  &  0.00  & -0.81   &  -0.10 \\
2.60 & 0.28261 & 0.28261 & 0.28459 &  0.28297  &  0.00  & -0.70  &   -0.13 \\

\hline
\end{tabular}
\end{center}
\end{table}
\end{center}

\begin{center}
\begin{table}[h]
\caption{Relative errors of the 4-loop approximant $\alpha^{(4)}_{ts}(Q^2,N_{ts})$, with $N_{tr}=4,6,9$. The approximant is compared with the exact numerical 4-loop coupling $\alpha_{num}^{(4)}$.}
\vspace{0.2cm}
\begin{center}

\begin{tabular}{lllll}
\hline
$\sqrt{Q^{2}}$  $ GeV $   &
$ \alpha_{num}^{(4)}$ &
Diff.(\%,ts)&
Diff.(\%,ts)&
Diff.(\%,ts) \\
&&$N_{tr}=4$&$N_{tr}=6$&$N_{tr}=9$\\ \hline

.80   &  .91262     &  15.7 & 9.4&  4.9  \\
.90   &  .69243     &   7.4 &3.3& 1.2  \\
1.00  &   .58701    &   4.3 & 1.6 & 0.4 \\
1.10  &   .52156    &   2.9 & 0.9 & 0.2  \\
1.20  &   .47585    &   2.1 & 0.5 & 0.1 \\
1.30  &   .44164    &   1.6 & 0.4 & 0.1  \\
1.40  &   .41484    &   1.2 & 0.3 & 0.0  \\
1.50  &   .39313    &   1.0 & 0.2 & 0.0  \\
1.60  &   .37510    &   0.8 & 0.1 & 0.0   \\
1.70  &   .35984    &   0.7 & 0.1 & 0.0   \\
1.80  &   .34670    &   0.6 & 0.1 & 0.0  \\
1.90  &   .33525    &   0.5 & 0.1 & 0.0  \\
2.00  &   .32515    &   0.5 & 0.1 & 0.0  \\

\hline
\end{tabular}
\end{center}
\end{table}
\end{center}

\begin{center}
\begin{table}[h]
\caption{The second order Euclidean expansion functions $\acal_{n}^{(2)}(Q^2)$ and $\acal_{as,n}^{(2)}(Q^2)$, for $n=1,2,$ as a function of the momentum transfer $\sqrt{Q^2}$. $\Lambda_{f=3}=0.400$ $GeV$.}
\vspace{0.2cm}
\begin{center}

\begin{tabular}{lllllll}
\hline
$\sqrt{Q^{2}}$  $ GeV $   &
$  {\acal}_{1}^{(2)}$ &
${\acal}_{as,1}^{(2)}$&
Diff.(\%,1)&
${\acal}_{2}^{(2)}$ &
 ${\acal}_{as,2}^{(2)}$&
Diff.(\%,2) \\ \hline
 0.4  & 0.507853  &  0.476767  &  6.1 & 0.118913    & 0.113524     & 4.5    \\
.60   & 0.438444  &  0.413090  &  5.8 & 0.107177    & 0.100299     & 6.4      \\
.80   & 0.393408  &  0.372388  &  5.3    & 0.09703     & 0.090135  & 7.1      \\
1.00  & 0.361380  &  0.343505 &  5.0   & 0.088705    & 0.082211  & 7.3     \\
1.20  & 0.337219  &  0.321662 &  4.6    & 0.081875    & 0.075872   & 7.3     \\
1.40  & 0.318218  &  0.304414 &  4.3    & 0.076208    & 0.070681   & 7.3     \\
1.60  & 0.302807  &  0.290363 &  4.1    & 0.071443    & 0.066345   & 7.1     \\
1.80  & 0.290003  &  0.278638 &    3.9    & 0.067385    & 0.062665  & 7.0    \\
2.00  & 0.279159  &   0.268669 &  3.8    & 0.063886    & 0.059498  & 6.9   \\
2.20  &  0.269831  &  0.260062 &  3.6    & 0.060839    & 0.05674   &  6.7     \\
2.40  &  0.261702  &  0.252536 &  3.5    & 0.05816    &  0.054315  & 6.6      \\
2.60  &  0.254538 &  0.245885 &   3.4   & 0.055784    & 0.052163   & 6.5    \\

\hline
\end{tabular}
\end{center}
\end{table}
\end{center}

\begin{center}
\begin{table}[h]
\caption{The second order Minkowskian expansion functions $\agoth_{n}^{(2)}(s)$ and $\agoth_{as,n}^{(2)}(s)$ (for $n=1,2$) as a function of the energy scale $\sqrt{s}$. $\Lambda_{f=3}=0.400$ $GeV$.}
\vspace{0.2cm}
\begin{center}

\begin{tabular}{lllllll}
\hline
$\sqrt{s}$  $ GeV $   &
$  {\agoth}_{1}^{(2)}$ &
${\agoth}_{as,1}^{(2)}$&
Diff.(\%,1)&
${\agoth}_{2}^{(2)}$ &
 ${\agoth}_{as,2}^{(2)}$&
Diff.(\%,2) \\ \hline
 0.4  & 0.495247   & 0.458395   & 7.44   &  0.133488 & 0.120209  & 9.9 \\
.60   & 0.417309   & 0.393376   & 5.73   &  0.114407 & 0.102211  & 10.7   \\
.80   & 0.369053   & 0.352141   & 4.58   &  0.098808 & 0.089398  & 9.5   \\
1.00  & 0.336208  &  0.323077   & 3.91   &  0.087040 & 0.079716  & 8.4  \\
1.20  & 0.312294  &  0.301367   & 3.50   &  0.078109 & 0.072190  & 7.6  \\
1.40  & 0.294001  &  0.284473   & 3.24   &  0.071173 & 0.066212  & 7.0  \\
1.60  & 0.279480  &  0.270906   & 3.07   &  0.065653 & 0 061368  & 6.5  \\
1.80  & 0.267616  &  0.259734   & 2.95   &  0.061160 & 0.057372  & 6.2   \\
2.00  & 0.257700  &  0.250344   & 2.85   &  0.057432 & 0.054022  & 5.9  \\
2.20  & 0.249258   & 0.242318   & 2.78   &  0.054286 & 0.051172  & 5.7    \\
2.40  & 0.241962   & 0.235360   & 2.73   &  0.051594 & 0.048718  & 5.6     \\
2.60  & 0.235575  &  0.229257   & 2.68  &   0.049261 & 0.046581    &5.4 \\

\hline
\end{tabular}
\end{center}
\end{table}
\end{center}

\section{The Euclidean and Minkowskian expansion functions}

For the reader's convenience here we briefly summarize main aspects of APT. Let $D(Q^2)$ be Adler D-function related to some timelike process. It has the asymptotic expansion in powers of a running coupling
\be
D_{pt}(Q^2)=D_{0}(1+\sum_{n=1}^{\infty}d_{n}\alpha_{s}^{n}(Q^2)),
\ee
here $D_0$ is a process dependent constant. In APT, the same quantity should be presented in the form of a nonpower asymptotic expansion as
\be
D_{an}(Q^2)=D_{0}(1+\sum_{n=1}^{\infty}d_n\acal_{n}(Q^2))
\ee
where ${\cal A}_n$ is the ``analyticized nth power'' of the coupling in the spacelike region. This is determined through the spectral representation
\be
\label{ann}
\acal_n(Q^2)=
\frac{1}{\pi}\int_{0}^{\infty}\frac{\rho_{n}(\sigma)}{\sigma+Q^2}d\sigma=
\frac{1}{\pi}\int_{-\infty}^{\infty}\frac{e^{t}}{(e^{t}+Q^2/{\Lambda}^2)}{\tilde
\rho_{n}(t)}dt,
\ee
where the spectral function is defined as
 $$\rho_{n}(\sigma)=\Im\{\alpha_{s}^{n}(-\sigma-\imath 0)\},$$
$\Lambda$ on the right of (\ref{ann}) denotes the QCD scale parameter and
$\tilde \rho_{n}(t)\equiv\rho_{n}(\sigma)$ with $t=\ln(\frac{\sigma}{\Lambda^2})$.
Let $R(s)$ be the physical quantity reconstructed through $D(Q^2)$ in the timelike domain (notable examples are $R_{e^{+}e^{-}}(s)=\sigma(e^{+}e^{-}\rightarrow hadrons)/\sigma(e^{+}e^{-}\rightarrow\mu^{+}\mu^{-})$ and $R_\tau(M_\tau)=\Gamma(\tau\rightarrow hadrons)/\Gamma(\tau\rightarrow \nu_{\tau}\nu_{\bar {e}}e)$.
Then, in APT it has the representation \cite{ss1}
\be
\label{gth}
\begin{array}{ll}
R(s)=R_0(1+r(s)) & \mbox{where}\quad r(s)=\sum_{n=1}^{\infty}d_n{\agoth}_{n}(s).
\end{array}
\ee
The timelike set of functions $\{\agoth_{n}(s)\}$ is defined by the formula \cite{ms1}
\be
\label{eleg}
\agoth_{n}(s)
=\frac{1}{\pi}\int_{s}^{\infty}\frac{d\sigma}{\sigma}\rho_n(\sigma).
\ee
We see, that the spectral functions corresponding to powers of the coupling are  of special importance. For $0<f\leq 6$, the spectral function corresponding to the nth power of the coupling (\ref{w2}) reads \cite{my1,mypr}
\be
\label{sf2}
{\tilde \rho}^{(2)}_{n}(t)=\left(\frac{\beta_{0}}{\beta_{1}}\right)^{n}\Im\left(-\frac{1}{1+W_{1}(z(t))}\right)^n;\quad z(t)=\frac{1}{b_{1}}\exp(-1-t/b_{1}+\imath \pi(1/b_{1}-1)),
\ee
one may rewrite the spectral function (\ref{sf2}) in the equivalent form, in terms of the branch $W_{-1}$, using the symmetry property $W_{k}(\bar z)=\overline {W_{-k}(z)}$ \cite{lamb}. Thus, we find
$$
{\tilde \rho}^{(2)}_{n}(t)=-\left(\frac{\beta_{0}}{\beta_{1}}\right)^{n}\Im\left(-\frac{1}{1+W_{-1}(z_{1}(t))}\right)^n,
$$
where $z_{1}(t)=\overline{z(t)}$.
The spectral function corresponding to the coupling (\ref{trs}) then is
\be
{\tilde \rho}_{ts,1}^{(k)}(t)=\sum_{n=1}^{N_{tr}}c_{n}^{(k)}{\tilde \rho}^{(2)}_{n}(t),
\ee
this gives the following representations for the first expansion functions
$$
\acal_{ts,1}^{(k)}(Q^2)=\sum_{n=1}^{n=N_{tr}}c_{n}^{(k)}\acal_{ts,n}^{(2)}(Q^2)
\quad
\mbox{and}\quad
\agoth_{ts,1}^{(k)}(s)=\sum_{n=1}^{n=N_{tr}}c_{n}^{(k)}\agoth_{ts,n}^{(2)}(s),
$$
analogical formulae hold for the functions with a higher index.

In practice integral (\ref{ann}) should be regulated. In Ref~\cite{mypr} the following formula was derived
\be
\label{rgl}
\acal_{n}(Q^2)=\acal_{n}(Q^2,R)+\agoth_{n}(\Lambda^{2}e^{R})+
\left\{\begin{array}{ll}
O({\bar Q}^2e^{-R}/R^{1+n}) &\mbox{if ${\bar Q}^2>1$}\\
O({\bar Q}^{-2}e^{-R}/R^{1+n}) &\mbox{if ${\bar Q}^2<1,$}
\end{array}
\right.
\ee
where $\acal_{n}(Q^2,R)$ denotes the integral
(\ref{ann}) taken over the finite interval $-R_{1}\leq t\leq R$, functions $\agoth_{n}$ , $n=1,2\ldots$, are defined in (\ref{eleg}) and ${\bar {Q}}^2=Q^2/\Lambda^2$.
For sufficiently large values of $R$, when ${\bar {Q}}^2\exp(-R)/R^{1+n}\ll 1$,
the contributions of order $e^{-R}$ can be omitted. So that the extra term $\agoth_{n}(\Lambda^{2}e^{R})$ compensates main error emerging because of truncation of the integral on the upper bound.
Formula (\ref{rgl}) enables us to achieve a good numerical precision even for moderate values of the cutoff $R$. In the region $t\rightarrow -\infty$ the integrand decreases rapidly, so that we can take $R_{1}< R$. To achieve a good numerical precision it suffices to limit the integral to the region
$-20<t<1000$.

To calculate the Minkowskian functions (\ref{eleg}) one may use a method of integrating expressions containing inverse functions \cite{lamb}. Using this technique, in work \cite{mypr} (see also papers \cite{join,howe}) the following expressions have been obtained
\begin{eqnarray}
\agoth_{1}^{(2)}(s)&=&-\frac{\beta_0}{\beta_1}+\frac{1}{\pi\beta_1}Im\left(\frac{1}{\alpha^{(2)}(-s)}\right)\nonumber\\
\agoth_{2}^{(2)}(s)&=&-\frac{1}{\pi\beta_1}Im\left(\ln\left(1+\frac{\beta_1}{\beta_0}\alpha^{(2)}(-s)\right)\right),
\end{eqnarray}
where
$$
\alpha^{(2)}(-s)\equiv\alpha^{(2)}(-s+\imath 0)=-\frac{\beta_{0}}{\beta_{1}}(1+W_{-1}(z_{s}))^{-1}\quad\mbox{with}\quad
z_{s}=b_{1}^{-1}(s{\Lambda}^{-2})^{-1/b_{1}}\exp(-1-\imath\pi(b_{1}^{-1}-1)).
$$
The Minkowskian functions with a higher index are easily determined by using the recursion formula
\be
\label{rf2}
\frac{\partial}{\partial \ln{s}}\agoth_{n}^{(2)}(s)=-n(\beta_{0}\agoth_{n+1}^{(2)}(s)+\beta_{1}\agoth_{n+2}^{(2)}(s))\quad n=1,2\ldots,
\ee
obtained in paper \cite{mypr}.

So far as the asymptotical solution (\ref{as}) is concerned one may use the following formula
\be
\label{usef}
L^{-n}\ln^{m} L=(-1)^{m}\frac{\partial}{\partial n^{m}}L^{-n},
\ee
where $L\equiv L(Q^2)=\ln (Q^2/\Lambda^2)$, $m$ and $n$ are positive numbers.
We can write, for $\sigma>0$,
\be
\label{phase}
L(-\sigma-\imath 0)=R_{\sigma}\exp{(\imath \Phi(\sigma))},\quad\mbox{where}\quad R_{\sigma}=\sqrt{\ln^{2}{\bar\sigma}+\pi^2}\quad \mbox{with}\quad \bar\sigma=\frac{\sigma}{\Lambda^2},
\ee
and
$$
\Phi(\sigma)=\left\{\begin{array}{ll}
-\arcsin{(\pi/R_{\sigma})}& \mbox{if $\bar\sigma>1,$}\\
-\pi+\arcsin{(\pi/R_{\sigma})}&\mbox{if $0<\bar\sigma <1.$}
\end{array}
\right.
$$
Evidently, the function $L^{-n}\ln^{m}{L}$ for $m,n>0$ satisfies the Kallen-Lehmann representation.  Let $\rho(\sigma,m,n)$ be the corresponding spectral density.
Combining (\ref{usef}) and (\ref{phase}) we readily derive the following formula
\be
\label{keyf}
\rho(\sigma,m,n)=(-1)^{m+1}\frac{\partial}{\partial n^{m}}\left(R_{\sigma}^{-n}\sin(n\Phi(\sigma))\right).
\ee
Formula (\ref{keyf}) allows one to derive explicit expressions
for the spectral functions corresponding to the asymptotical coupling (\ref{as}) (and to its positive powers) in various orders of perturbation theory. Thus, we have
\be
\label{asspec}
\begin{array}{l}
\displaystyle{\rho_{as,1}^{(1)}(\sigma)=\frac{\pi}{\beta_{0}(\ln^{2}\bar {\sigma}+\pi^2)}},\\
\displaystyle{\rho_{as,1}^{(2)}(\sigma)=\rho_{as,1}^{(1)}(\sigma)-\frac{\beta_{1}}{\beta_{0}^{3}}\rho(\sigma,1,2)=}\\
\displaystyle{\frac{\pi}{\beta_{0}(\ln^{2}\bar {\sigma}+\pi^2)}-\frac{\beta_{1}}{\beta_{0}^{3}}\frac{1}{(\ln^{2}\bar {\sigma}+\pi^2)^2}(\pi\ln{\bar\sigma}\ln{(\ln^{2}\bar{\sigma}+\pi^2)}+(\ln^{2}\bar{\sigma}-\pi^2)\Phi(\sigma))},\\
\displaystyle{\rho_{as,1}^{(3)}(\sigma)=\rho_{as,1}^{(2)}(\sigma)+\frac{1}{\beta_0^3}\left(\frac{\beta_1^2}{\beta_0^2}(\rho(\sigma,2,3)-\rho(\sigma,1,3))+\left(\frac{\beta_2}{\beta_0}-\frac{\beta_1^2}{\beta_0^2}\right)\rho(\sigma,0,3)\right), \quad \mbox{etc.}}
\end{array}
\ee
To reconstruct the timelike analytic images of the powers of the coupling (\ref{as}) we define the auxiliary quantities
\be
\label{ast}
\agoth(s,m,n)=\frac{1}{\pi}\int_{s}^{\infty}\rho(\sigma,m,n)d\sigma.
\ee
In the case $s>\Lambda^2$, one may change the integration variable in (\ref{ast}) according $u=\arcsin(\pi/R_{\sigma})$. We thus find the  representation
\be
\label{ast2}
\agoth(s,m,n)=(-1)^{m}\frac{\partial}{\partial n^{m}}\left(\pi^{-n}\int_{0}^{\arcsin{(\pi/R_{s})}}\sin^{n-2}{u}\sin(n u)du\right),
\ee
here use has been made of (\ref{keyf}). For $0<s<\Lambda^2$, the analogical formula can be written.
In the most cases of practical interest the integrals (\ref{ast2}) can be performed in terms of elementary functions. Using (\ref{ast2}),
for example,
in the 2-loop case, if $\bar s=s/\Lambda^2>1$, we find
\be
\agoth_{as,1}^{(2)}(s)=\frac{1}{\beta_0}\left(0.5-\frac{1}{\pi}\arctan\left(\frac{\ln {\bar s}}{\pi}\right)\right)+\frac{\beta_1}{\beta_0^3}\frac{1}{R_{s}^2}\left(-1-\ln R_{s}+\frac{\ln {\bar s}}{\pi}\arcsin\left(\frac{\pi}{R_{s}}\right)\right),
\ee
where $R_s=\sqrt{\ln^2 {\bar s}+\pi^2}$ and $\bar{s}=s/\Lambda^2$

In Table 4 we present numerical results obtained with the Euclidean expansion functions in the 2-loop case.
$\acal_{n}^{(2)}(Q^2)$ stands for the  Euclidean analytic image of the nth power of the exact explicit 2-loop coupling (\ref{w2}).
The Euclidean analytic images of
the powers of the 2-loop asymptotical coupling (the first line in  Eq.~(\ref{as})) are denoted as
$\acal_{as,n}^{(2)}(Q^2)$, $n=1,2\ldots$. We examine the first two Euclidean functions $(n=1,2)$ in the region with three flavours 0.4 $GeV$ $<\sqrt{Q^2}<$2.6 $GeV$. We see, that the errors in the approximations $\acal_{as,n}^{(2)}(Q^2)$ for $n=1,2$ are large.
The relative errors are determined as Diff.(\%,n)$=(1-\acal _{as,n}^{(2)}(Q^2)/\acal_{n}^{(2)}(Q^2))*100$.

In Table 5 we give the 2-loop order results obtained with the corresponding Minkowskian expansion functions $\agoth_{n}^{(2)}(s) $ and $\agoth_{as,n}^{(2)}(s)$ for $n=1,2$ in the region 0.4 $GeV$ $<\sqrt{s}<2.6$ $GeV$. It is seen from the table  that the errors (Diff.(\%,n)$=(1-\agoth_{as,n}^{(2)}(s)/\agoth_{n}^{(2)}(s))*100$) in the asymptotical approximants $\agoth_{as,n}^{(2)}(s)$ are too large.

In Tables 6 and 7 we estimate the accuracy of the 3-loop order approximants
$\acal_{ts,n}^{(3)}(Q^2)$ and $\agoth_{ts,n}^{(3)}(s)$ that are constructed from the truncated series (\ref{trs}) where $N_{tr}=5$. In the domain with three flavours, we compare these functions with the ``exact'' functions $\acal_{num,n}^{(3)}(Q^2)$ and $\agoth_{num,n}^{(3)}(s)$. The ``exact'' functions are reconstructed numerically by solving the transcendental equation
(\ref{3lim}) in the complex-$Q^2$ plane.
We see, that both (Euclidean and Minkowskian) Lambert-W approximants give the excellent accuracy in the considered region. Beyond this region, these approximations are even more accurate.

In Table 8, we summarize the numerical results obtained with the 3-loop Euclidean functions $\acal_{ts,n}^{(3)}(Q^2)$ and $\acal_{as,n}^{(3)}(Q^2)$ for $n=1,2$. We take again $N_{tr}=5$. The functions $\acal_{as,n}^{(3)}(Q^2)$ correspond to the 3-loop asymptotical coupling. We see that the theoretical errors (Diff.(\%,n)$=(1-\acal_{as,n}^{(3)}/\acal_{ts,n}^{(3)})*100$) in $\acal_{as,n}^{(3)}(Q^2)$ for $n=1,2$ are significantly smaller than those in the 2-loop case.
The 3-loop results for the Minkowskian expansion functions are given in Table 9. We see, that the errors in the approximants $\agoth_{as,n}^{(3)}(s)$ are approximately of the same size as those in the corresponding 3-loop Euclidean expansion functions.

In Tables 10 and 11, the results obtained in the 4-loop case for the Euclidean and Minkowskian expansion functions are given. $\acal_{num,1}^{(4)}(Q^2)$ and $\agoth_{num,1}^{(4)}(s)$ stand for the ``exact'' numerical functions obtained by solving the transcendental equation (\ref{gsol}) (to fourth order) in the complex $Q^2$-plane (see Eq.~(\ref{4lim})). The functions $\acal_{ts,1}^{(4)}(Q^2)$ and $\agoth_{ts,1}^{(4)}(s)$ are reconstructed from the truncated series (\ref{trs}) where $N_{tr}=6$. The approximants $\acal_{as,1}^{(4)}(Q^2)$ and $\agoth_{as,1}^{(4)}(s)$ correspond to the asymptotical coupling (\ref{as}). We reproduce practically exact results with the approximants $\acal_{ts,1}^{(4)}(Q^2)$ and $\agoth_{ts,1}^{(4)}(s)$. The asymptotical approximants $\acal_{as,1}^{(4)}(Q^2)$ and $\agoth_{as,1}^{(4)}(s)$ lead to the results, also, in fairly good agreement with the exact functions.

\section{Conclusion}

The explicit Lambert-W solutions and the conventional asymptotical approximations to the QCD RG equation are compared with one another in the $\overline {MS}$ scheme up to 4-loop order.
These approximations have been carefully examined in the infrared region, where they are expected to be less accurate. As a standard for the comparison we have used the ``exact'' numerically calculated running coupling.
Beyond second order, the Lambert-W approximations were represented as series in powers of the exact 2-loop coupling. We have shown, that the partial sums of  these series, with the first few terms, give sufficiently good successive approximations to the ``exact'' numerical coupling. The required accuracy is achieved by choice of a sufficiently large value for the truncation order $N_{tr}$, even close to the Landau singularity (see Figure 1). This was confirmed to third, as well to fourth orders (see Tables 1 and 2)\footnote{We shall give a mathematical investigation of the convergence properties of the series (\ref{ser}) elsewhere.}. Note that, at the energy scale of the $\tau$-lepton mass $M_{\tau}=1.77$ $GeV$, for $\Lambda=0.400$ $GeV$, the error in $\alpha^{(4)}_{ts}(Q^2, N_{ts}=9)$ is about 0.01\%, while the error in $\alpha_{as}^{(4)}(Q^2)$, at the same scale, is more sizable; Diff.(\%,as)=1.6.

Applications of these solutions in APT are discussed.
We have presented convenient theoretical expressions for the Euclidean and Minkowskian analytic images of powers of the running coupling, which are determined in terms of the Lambert-W function (see sect.3). Alternative expressions have been also derived using asymptotical solution (\ref{as}). In the 2-loop case, the asymptotical images (i.e. the images of the asymptotical coupling (\ref{as})) are found to lead to large theoretical errors. For example, the errors in the functions $\agoth_{as,1}^{(2)}(m_{\rho}^2)$ and $\agoth_{as,1}^{(2)}(M_{\tau}^2)$ ($m_{\rho}=0.770$ $GeV$ and $M_{\tau}=1.77$ $GeV$ being masses of the $\rho$ mezon and $\tau$-lepton) are found to be 4.6\% and 3\% respectively.
However, we have observed, to third and fourth orders, that the accuracy of the asymptotical Euclidean and Minkowskian approximants is significantly improved. Thus, the errors in
$\agoth_{as,1}^{(3)}(s)$, for $s=m_{\rho}^2$ and $s=M_{\tau}^2$, are found to be 0.9\% and 0.6\% respectively (see Table 9). In the 4-loop case, these errors are even smaller; 0.4\% and 0.5\% respectively (Table 11).

On the other hand, we have reproduced practically exact results with the Euclidean and Minkowskian images for the Lambert-W coupling (\ref{trs}) to third and fourth orders (see Tables 8,9,10 and 11). Another argument in favor to the Lambert-W solutions is that they enable one to derive closed analytic expressions for the time-like observables \cite{join,howe}.
Of considerable importance in numerical and analytical calculations are the recursion formulas (\ref{rec4}) and (\ref{rf2}) which enable one to write compact and clear Maple programs.

\begin{center}
\begin{table}[h]
\caption{ The Euclidean approximant $\acal_{ts,1}^{(3)}(Q^2)$ reconstructed from the truncated series (\ref{trs}) with $N_{tr}=5$ vs the ``exact'' (numerical) 3-loop function $\acal_{num,1}^{(3)}(Q^2)$. $\Lambda_{f=3}=0.400$ $GeV$.}
\vspace{0.2cm}
\begin{center}

\begin{tabular}{llll||llll}
\hline
$\sqrt{Q^2}$  $ GeV $   &
$ \acal_{num,1}^{(3)}$ &
$ \acal_{ts,1}^{(3)}$ &
Diff.(\%) &
$\sqrt{Q^2}$  $ GeV $   &
$ \acal_{num,1}^{(3)}$ &
$ \acal_{ts,1}^{(3)}$ &
Diff.(\%) \\ \hline

 0.4 & 0.511938 & 0.511933 & 0.001 & 1.8& 0.295068 & 0.295069 & -0.000 \\
 0.6 & 0.443765 & 0.443761 & 0.001 & 2.0& 0.284026 & 0.284028 & -0.001 \\
 0.8 & 0.399121 & 0.399119 & 0.000 & 2.2& 0.274510 & 0.274511 & -0.000 \\
 1.0 & 0.367134 & 0.367133 & 0.000 & 2.4& 0.266203 & 0.266205 & -0.001\\
 1.5 & 0.315504 & 0.315505 & -0.000 & 2.6& 0.258874 & 0.258876 & -0.001 \\
\hline
\end{tabular}
\end{center}
\end{table}
\end{center}

\begin{center}
\begin{table}[h]
\caption{ The Minkowskian approximant $\agoth_{ts,1}^{(3)}(s)$ corresponding to the truncated series (\ref{trs}) where $N_{tr}=5$ vs the ``exact'' (numerical) 3-loop function $\agoth_{num,1}^{(3)}$. $\Lambda_{f=3}=0.400$ $GeV$.}
\vspace{0.2cm}
\begin{center}

\begin{tabular}{llll||llll}
\hline
$\sqrt{s}$  $ GeV $   &
$ \agoth_{num,1}^{(3)}$ &
$ \agoth_{ts,1}^{(3)}$ &
Diff(\%) &
$\sqrt{s}$  $ GeV $   &
$ \agoth_{num,1}^{(3)}$ &
$ \agoth_{ts,1}^{(3)}$ &
Diff(\%) \\ \hline

 0.4 & 0.501635 & 0.501609 &0.005  & 1.8& 0.273008 &.0.273011 & -0.001 \\
 0.6 & 0.425402 & 0.425397 & 0.001 & 2.0& 0.262718 & 0.262720 & -0.001 \\
 0.8 & 0.377104 & 0.377110 & -0.002& 2.2& 0.253954 & 0.253955 & -0.000 \\
 1.0 & 0.343725 & 0.343733 & -0.002& 2.4& 0.246377 & 0.246378 & -0.000 \\
 1.5 & 0.292420 & 0.292425 & -0.002& 2.6& 0.239746 & 0.239747 & -0.000 \\

\hline
\end{tabular}
\end{center}
\end{table}
\end{center}

\begin{center}
\begin{table}[h]
\caption{ The third order Euclidean approximants $\acal_{ts,n}^{(3)}(Q^2)$ and $\acal_{as,n}^{(3)}(Q^2)$ (n=1,2) as a function of the momentum transfer $\sqrt{Q^2}$. Diff.(\%,n)$=(1-\acal_{as,n}^{(3)}/\acal_{ts,n}^{(3)})*100$.}
\vspace{0.2cm}
\begin{center}

\begin{tabular}{lllllll}

$\sqrt{Q^{2}}$  $ GeV $  &
$  {\acal}_{ts,1}^{(3)}$ &
$  {\acal}_{as,1}^{(3)}$ &
Diff.(\%,1)&
${\acal}_{ts,2}^{(3)}$ &
$  {\acal}_{as,2}^{(3)}$ &
Diff.(\%,2) \\ \hline

0.10   & 0.754824   & 0.754546  & 0.04  & 0.118112  & 0.115945  & 1.83 \\
0.20   & 0.635380   & 0.638860  &-0.55  & 0.124962  & 0.123223  & 1.39 \\
0.30   & 0.562731   & 0.567327  &-0.82  & 0.122311  & 0.121454  & 0.70  \\
0.40   & 0.511933   & 0.516651  &-0.92  & 0.117273  & 0.117054  & 0.19   \\
0.50   & 0.473770   & 0.478275  &-0.95  & 0.111773  & 0.111952  & -0.16    \\
0.60   & 0.443761   & 0.447960  &-0.95  & 0.106436  & 0.106854  & -0.39     \\
0.70   & 0.419392   & 0.423279  &-0.93  & 0.101465  & 0.102025  & -0.55     \\
0.80   & 0.399119   & 0.402718  &-0.90  & 0.096911  & 0.097551  & -0.66    \\
0.90   & 0.381931   & 0.385274  &-0.88  & 0.092763  & 0.093446  & -0.74    \\
1.00  &  0.367133   & 0.370252  &-0.85  & 0.088990  & 0.089694  & -0.79    \\
1.20  & 0.342862   & 0.345611  &-0.80  & 0.082421  & 0.083127  & -0.86    \\
1.40  & 0.323686   & 0.326151  &-0.76  & 0.076922  & 0.077607  & -0.89    \\
1.60  & 0.308076   & 0.310316  &-0.73  & 0.072264  & 0.072920  & -0.91      \\
1.80  & 0.295069   & 0.297128  &-0.70  & 0.068273  & 0.068898  & -0.92      \\
2.00  & 0.284028   & 0.285938  &-0.67  & 0.064815  & 0.065410  & -0.92      \\
2.20  & 0.274511   & 0.276297  &-0.65  & 0.061790  & 0.062356  & -0.92      \\
2.40  & 0.266205   & 0.267885  &-0.63  & 0.059121  & 0.059661  & -0.91      \\
2.60  & 0.258876   & 0.260465  &-0.61  & 0.056747  &.057262    & -0.91      \\

\hline
\end{tabular}
\end{center}
\end{table}
\end{center}

\begin{center}
\begin{table}[h]
\caption{The third order Minkowskian approximants $\agoth_{ts,n}^{(3)}(s)$ and $\agoth_{as,n}^{(3)}(s)$ (n=1,2), in the domain with three flavours. Diff(\%,n)$=(1-\agoth_{as,n}^{(3)}/\agoth_{ts,n}^{(3)})*100$.}
\vspace{0.2cm}
\begin{center}

\begin{tabular}{lllllll}
$\sqrt{s}$  $ GeV $   &
$  {\agoth}_{ts,1}^{(3)}$ &
$  {\agoth}_{as,1}^{(3)}$ &
Diff.(\%,1)&
${\agoth}_{ts,2}^{(3)}$ &
$ {\agoth}_{as,2}^{(3)}$ &
Diff.(\%,2)
 \\ \hline

0.10  & 0.769627     & 0.767795  & 0.24 &  0.126247   &   0.122167 & 3.23 \\
0.20  & 0.640566     & 0.648163  &-1.19 & 0.140034  &   0.136232   & 2.71   \\
0.30  & 0.559089     & 0.568235  &-1.64 & 0.138017  &   0.137921   &  0.07   \\
0.40  & 0.501609     & 0.508946  &-1.46 & 0.130889  &   0.132644   & -1.34  \\
0.50  & 0.458689     & 0.464136  &-1.19 & 0.122465  &   0.124524   & -1.68   \\
0.60  & 0.425397     & 0.429557  &-0.98 & 0.114203   &   0.116042   & -1.61  \\
0.70  & 0.398821     & 0.402184  &-0.84 & 0.106619   &   0.108168   & -1.45  \\
0.80  & 0.377110     & 0.379979  &-0.76 & 0.099846   &   0.101157   & -1.31   \\
0.90  & 0.359032     & 0.361579  &-0.71 & 0.093863   &   0.094997   &-1.21    \\
1.00 & 0.343733    &  0.346057 &-0.68 &  0.088593   &   0.089597  & -1.13    \\
1.20 & 0.319204    &  0.321235 &-0.64 &  0.079832   &   0.080667  & -1.05    \\
1.40 & 0.300337    &  0.302175 &-0.61 &  0.072913   &   0.073643  & -1.00    \\
1.60 & 0.285311    &  0.287003 &-0.59 &  0.067343   &   0.067999  & -0.97  \\
1.80 & 0.273011    &  0.274587 &-0.58 &  0.062776   &   0.063374  & -0.95   \\
2.00 & 0.262720    &  0.264199 &-0.56 &  0.058965   &   0.059517  & -0.94    \\
2.20 & 0.253955    &  0.255351 &-0.55 &  0.055737   &   0.056251  & -0.92    \\
2.40 & 0.246378    &  0.247702 &-0.54 &  0.052967   &   0.053448  & -0.91    \\
2.60 & 0.239747    &  0.241007 &-0.53 &  0.050561   &   0.051014  & -0.89    \\
\hline
\end{tabular}
\end{center}
\end{table}
\end{center}

\begin{center}
\begin{table}[h]
\caption{ The fourth order Euclidean approximants $\acal_{ts,1}^{(4)}(Q^2, N_{ts}=6)$ and $\acal_{as,1}^{(4)}(Q^2)$ vs the ``exact'' numerical function $\acal_{num,1}^{(4)}(Q^2)$. $\Lambda_{f=3}=0.400$ $GeV$.}
\vspace{0.2cm}
\begin{center}

\begin{tabular}{cccccc}
$\sqrt{Q^{2}}$  $ GeV $   &
$ \acal_{num,1}^{(4)}$ &
$\acal_{ts,1}^{(4)} $ &
$\acal_{as,1}^{(4)}$ &
Diff.(\%,ts) &
Diff.(\%,as) \\
\hline

0.1  & 0.754123 & 0.754128 &  0.755096  &- 0.001  &  -0.13  \\
0.2  & 0.634500 & 0.634501 &  0.632852  & -0.000  & 0.26  \\
0.3  & 0.561976 & 0.561972 &  0.559131  &0.001  & 0.51  \\
0.4  & 0.511361 & 0.511355 &  0.508229  &0.001  & 0.61  \\
0.5  & 0.473372 & 0.473365 &  0.470371  &0.001  & 0.63  \\
0.6  & 0.443510 & 0.443504 &  0.440810  &0.001  & 0.61  \\
0.7  & 0.419262 & 0.419257 &  0.416912  &0.001  & 0.56   \\
0.8  & 0.399087 & 0.399083 &  0.397085  &0.001  & 0.50   \\
0.9  & 0.381977 & 0.381975 &  0.380298  &0.000  & 0.44   \\
1.0  & 0.367243 & 0.367242 &  0.365854  &0.000  & 0.38   \\
1.2  & 0.343063 & 0.343063 &  0.342154  & -0.000  & 0.27   \\
1.4  & 0.323946 & 0.323948 &  0.323404  & -0.001  & 0.17   \\
1.6  & 0.308374 & 0.308376 &  0.308108  & -0.001  & 0.09   \\
1.8  & 0.295391 & 0.295394 &  0.295334  & -0.001  & 0.02   \\
2.0  & 0.284365 & 0.284367 &  0.284466  & -0.001  &  -0.04   \\
2.2  & 0.274857 & 0.274859 &  0.275080  & -0.001  &  -0.08   \\
2.4  & 0.266554 & 0.266556 &  0.266870  & -0.001  &  -0.12    \\
2.6  & 0.259226 & 0.259228 &  0.259613  & -0.001  &  -0.15    \\
\hline
\end{tabular}
\end{center}
\end{table}
\end{center}

\begin{center}
\begin{table}[h]
\caption{The fourth order Minkowskian approximants $\agoth_{ts,1}^{(4)}(s, N_{ts}=6)$ and $\agoth_{as,1}^{(4)}(s)$ vs the ``exact'' numerical function $\agoth_{num,1}^{(4)}(s)$. $\Lambda_{f=3}=0.400$ $GeV$.}
\vspace{0.2cm}
\begin{center}

\begin{tabular}{cccccc}
\hline
$\sqrt{s}$  $ GeV $   &
$ \agoth_{num,1}^{(4)}$ &
$ \agoth_{ts,1}^{(4)}$&
$ \agoth_{as,1}^{(4)}$&
Diff.(\%,ts)&
Diff.(\%,as) \\ \hline

 0.1  & 0.768748 &  0.768760  &  0.772631 & -0.002 &   -0.51   \\
 0.2  & 0.639012 &  0.639034  &  0.635957 & -0.003 &  0.48  \\
 0.3  & 0.557764 &  0.557745  &  0.550459 & -0.003  & 1.31   \\
 0.4  & 0.500760 &  0.500717  &  0.493308 & -0.009  & 1.49  \\
 0.5  & 0.458283 &  0.458247  &  0.452271 & -0.008  & 1.31  \\
 0.6  & 0.425335 &  0.425315  &  0.421015 & -0.005  & 1.02  \\
 0.7  & 0.399005 &  0.399001  &  0.396164 & -0.001  &  0.71   \\
 0.8  & 0.377464 &  0.377470  &  0.375788 & -0.002  &  0.44   \\
 0.9  & 0.359498 &  0.359509  &  0.358695 & -0.003  &  0.22    \\
 1.0  & 0.344270 &  0.344284  &  0.344107 & -0.004  &  0.05    \\
 1.2  & 0.319808 &  0.319822  &  0.320436 & -0.004  &  -0.20      \\
 1.4  & 0.300954 &  0.300966  &  0.301979 & -0.004  &  -0.34      \\
 1.6  & 0.285916 &  0.285925  &  0.287128 & -0.003  &  -0.42      \\
 1.8  & 0.273593 &  0.273600  &  0.274881 & -0.003  &  -0.47     \\
 2.0  & 0.263275 &  0.263281  &  0.264579 & -0.002  &  -0.50      \\
 2.2  & 0.254483 &  0.254487  &  0.255770 & -0.002  &  -0.51      \\
 2.4 & 0.246879 &   0.246882  &  0.248133 & -0.001  &  -0.51       \\
 2.6 & 0.240222 &   0.240224  &  0.241433 & -0.001  &  -0.50        \\

\hline
\end{tabular}
\end{center}
\end{table}
\end{center}

\end{document}